\documentclass[prd,aps,nofootinbib,showpacs,10pt]{revtex4}
\usepackage{amsmath,graphicx,epsfig,amssymb,dsfont,mathtools}

\usepackage[usenames]{color}
\usepackage{ulem} 
\usepackage{bigstrut}

\begin{document} 
\title{To understand the rare decay $B_s\to\pi^+\pi^-\ell^+\ell^-$}
\author{ Wei Wang and Rui-Lin Zhu}

\affiliation{ 
$^1$ INPAC, Shanghai Key Laboratory for Particle Physics and Cosmology, Department of Physics and Astronomy, Shanghai Jiao-Tong University, Shanghai, 200240,   China\\
$^2$
State Key Laboratory of Theoretical Physics, Institute of Theoretical Physics, Chinese Academy of Sciences, Beijing 100190, China}

\begin{abstract}
Motivated by the LHCb measurement, we analyze the $B_s\to \pi^+\pi^-\ell^+\ell^-$ decay in the kinematics  region where the pion pairs  have invariant mass in the range $0.5$-$1.3$ GeV and muon pairs do not originate from a  resonance. The scalar $\pi^+\pi^-$ form factor induced by the strange $\bar ss$ current is predicted by the unitarized approach rooted in the chiral perturbation theory. Using the two-hadron light-cone distribution amplitude,   we then can derive  the $B_s\to \pi^+\pi^-$ transition form factor in  the light-cone sum rules approach.   Merging these quantities, we  present  our results  for differential decay width which can generally  agree with the experimental data.  More accurate measurements at the LHC and KEKB  in future are helpful  to  validate our formalism and determine the inputs in this approach. 
\end{abstract}
\maketitle



\begin{figure}\begin{center} 
\includegraphics[scale=0.5]{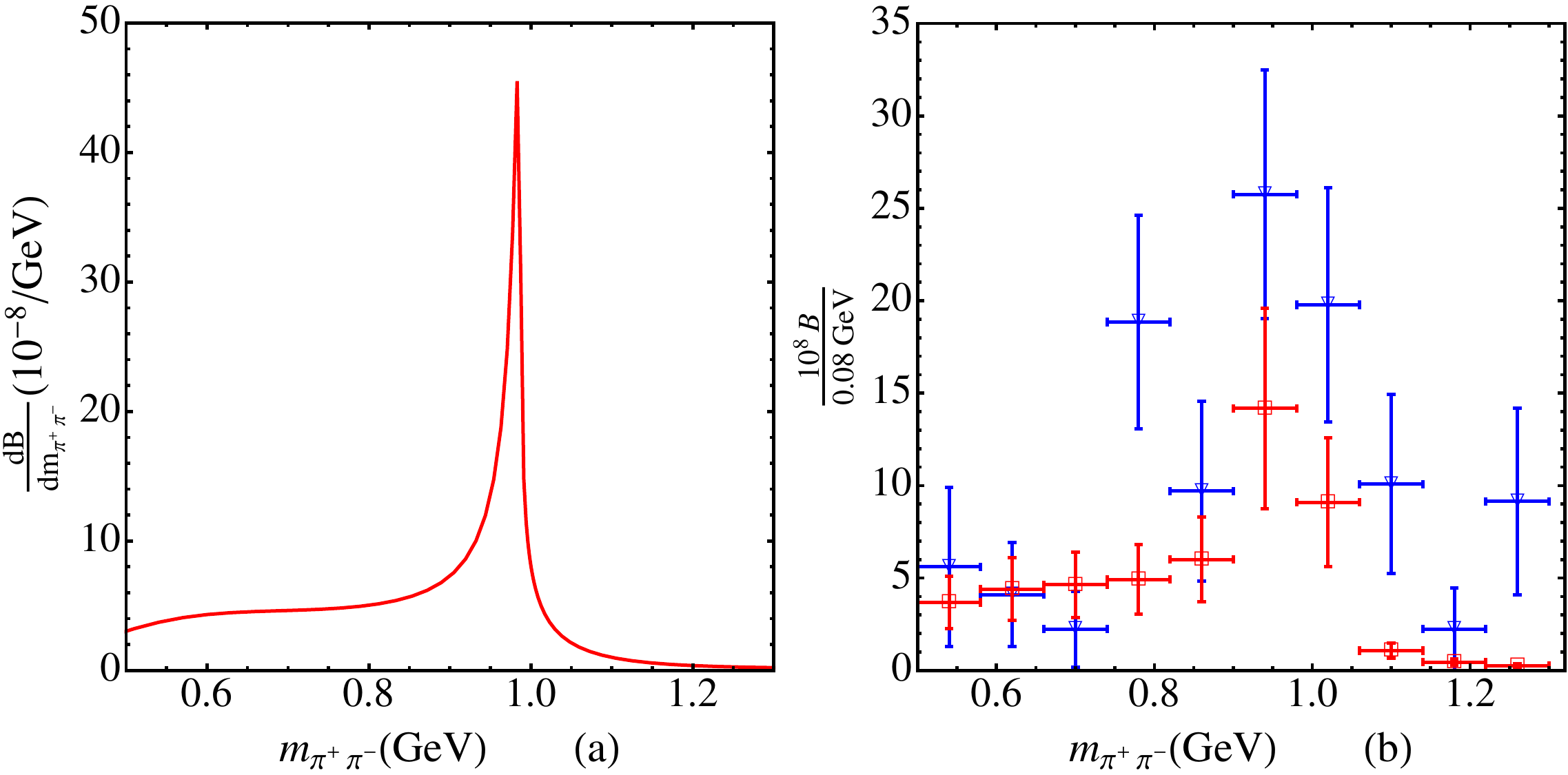} 
\caption{The differential branching ratio  for the $B_s\to \pi^+\pi^-\ell^+\ell^-$.  The experimental data (with triangle markers)  has been normalized to the central value of the branching fraction: ${\cal B}(B_s^0\to \pi^+\pi^-\mu^+\mu^-)=(8.6\pm 1.5\pm 0.7\pm 0.7)\times 10^{-8}$. Theoretical predictions (with square markers) are based on the result for the time-like scalar form factors  derived in  the unitarized CHPT. } \label{fig:dBRdM2h}
\end{center}
\end{figure}

 
Very recently, the LHCb has performed an analysis of rare $B_s$ decays into the $\pi^+\pi^-\mu^+\mu^-$ final state~\cite{Aaij:2014lba}  and the branching fraction  is measured as
\begin{eqnarray}
{\cal B}(B_s\to  \pi^+\pi^- \mu^+\mu^-) = (8.6\pm 1.5\pm 0.7\pm 0.7)\times 10^{-8}, \label{eq:Bspipidata}
\end{eqnarray}
where the first two errors are statistical,  and systematic respectively.  The third error is due to uncertainties on the normalization, i.e. the branching fraction of the  $B^0\to J/\psi(\to \mu^+\mu^-) K^*(\to K^+\pi^-)$.     The branching fraction for  $B_s\to f_0(980) \mu^+\mu^-$~\cite{Aaij:2014lba}  is determined as: 
\begin{eqnarray}
{\cal B}(B_s\to f_0(980)(\to \pi^+\pi^-) \mu^+\mu^-) = (8.3\pm 1.7)\times 10^{-8}, \label{eq:Bs_f0_mumu_data}
\end{eqnarray} 
which lies in the vicinity of the total branching fraction in Eq.~\eqref{eq:Bspipidata}. Despite the errors, the closeness of the two branching fractions and the differential distribution as shown later in Fig.~(\ref{fig:dBRdM2h}b) may indicate  the dominance of the $f_0(980)$ contributions in the $B_s\to \pi^+\pi^-\mu^+\mu^-$.

The $B_s\to \pi^+\pi^-\mu^+\mu^-$ is a four-body process.  
Its decay amplitude   shows two distinctive features.  On the one side, the $\pi^+\pi^-$ final state interaction is constrained by unitarity and analyticity. On the other side, the  $b$ mass scale is much higher than the hadronic scale $\Lambda_{\rm QCD}$, which allows an expansion of  the hard-scattering kernels in terms of the strong coupling constant $\alpha_s$ and the dimensionless power-scaling  parameter $\Lambda_{\rm QCD}/m_b$.  
In Refs.~\cite{Meissner:2013hya,Meissner:2013pba,Doring:2013wka}, we have  developed a formalism that makes use of   these two advantages. This approach  was also pioneered in Ref.~\cite{Gardner:2001gc,Maul:2001zn}, and see also Refs.~\cite{Chen:2002th,Chen:2004az,Wang:2014ira,Wang:2014qya} for applications to charmless three-body $B$ decays.  In doing this,  the new formalism can simultaneously  merge the perturbation theory  at the $m_b$ scale and the low-energy effective theory based on  the chiral symmetry to describe the S-wave $\pi\pi$ scattering. The aim of this work is to further examine this formalism by confronting this theoretical framework with  the recent  data on  $B_s\to \pi^+\pi^-\mu^+\mu^-$.    An independent analysis that is based on the perturbative QCD approach is also under progress~\cite{Wang:2015uea}


We   start with the  differential decay width  for $B_s\to \pi^+\pi^-\ell^+\ell^-$.  The effective Hamiltonian for the transition  $b\to s\ell^+\ell^-$  
 \begin{eqnarray}
 {\cal
 H}_{\rm{eff}}=
 -\frac{G_F}{\sqrt{2}}V_{tb}V^*_{ts}\sum_{i=1}^{10}C_i(\mu)O_i(\mu)
 \nonumber\label{eq:Hamiltonian}
 \end{eqnarray}
involves various  four-quark and the magnetic penguin operators $O_i$.  the $C_i(\mu)$
are the corresponding  Wilson coefficients for these local operators $O_i$. 
$G_F$ is the Fermi constant, and $V_{tb}=0.99914\pm0.00005$ and
$V_{ts}=-0.0405^{+0.011}_{-0.012}$~\cite{Agashe:2014kda} are the CKM matrix elements.  
The $b$ and $s$ quark masses are $m_b=(4.66\pm 0.03)$GeV and
$m_s=(0.095\pm0.005)$GeV~\cite{Agashe:2014kda}.  The  $b\to sl^+l^-$  transition has the decay amplitude  
\begin{eqnarray}
 i {\cal M}(b\to
 s\ell^+\ell^-)&=& iN_1\times
 \bigg\{ ({C_9+C_{10}})[\bar sb]_{V-A}[\bar \ell\ell]_{V+A}
 +({C_9-C_{10}})[\bar sb]_{V-A}[\bar \ell\ell]_{V-A}  \nonumber\\
 && + 4C_{7L}m_b[\bar s i\sigma_{\mu\nu}
 (1+\gamma_5)b]\frac{q^\mu}{q^2}\times[\bar \ell\gamma^\nu \ell]
 + 4C_{7R}m_b[\bar s i\sigma_{\mu\nu}
 (1-\gamma_5)b]\frac{q^\mu}{q^2}\times[\bar \ell \gamma^\nu
 \ell]\bigg\},\label{eq:decay-amplitude-bsll-LR}
\end{eqnarray}
where $C_{7L}=C_7$ and $C_{7R}=C_{7L}{m_s}/{m_b}$, and 
\begin{eqnarray}
N_1= \frac{ G_F} {4\sqrt 2} \frac{\alpha_{\rm em}}{\pi} V_{tb}V_{ts}^*. 
\end{eqnarray}

The   $B\to M_1M_2\ell^+\ell^-$ is    a four-body decay mode, whose decay amplitude can be obtained by sandwiching
Eq.~\eqref{eq:decay-amplitude-bsll-LR} between the initial and final hadronic states. The spinor product $[\bar sb]$  will be replaced by corresponding  hadronic matrix elements.  A   general  differential decay width for  $B\to M_1M_2\ell^+\ell^-$ with various partial wave contributions  has been derived using the helicity amplitude in Ref.~\cite{Lu:2011jm}.  In the $B_s\to \pi^+\pi^-\mu^+\mu^-$ case, the S-wave contribution will dominate and thus the angular distribution  is  derived as 
\begin{eqnarray}
 \frac{d^3\Gamma}{dm_{\pi\pi}^2dq^2  d\cos\theta_l  }
 &=& \frac{3}{8}\Big[J_1^c   + J_2^c  
 \cos(2\theta_l)  \Big],
\end{eqnarray} 
where 
$\theta_l$  is the polar angle  between the  $\mu^-$ and the $B_s$ moving direction  in the  lepton pair rest frame. The angular coefficients  are given by 
\begin{eqnarray}
J_1^c&=&   \bigg\{   |{\cal A}^0_{L0}|^2+|{\cal A}^0_{R0}|^2
 +8  \hat m_l^2  | {\cal A}^0_{L0}{\cal A}^{0*}_{R0} | \cos(\delta_{L0}^0 -\delta_{R0}^0) 
 +4 \hat m_l^2  |{\cal A}_t^0|^2 \bigg\},  \\  
 J_2^c  &=& -\beta_l^2   \bigg\{   |{\cal A}^0_{L0}|^2+|{\cal A}^0_{R0}|^2    \bigg\} . 
  \label{eq:simplified_angularCoefficients_S-wave}
\end{eqnarray} 
In the above equations, $\beta_\ell=
\sqrt{1-4m_\ell^2/q^2}$, and $\hat m_\ell= m_\ell/\sqrt{q^2}$. 
The helicity amplitude is 
\begin{eqnarray}
 {\cal A}_{L/R,0}^0&=&\sqrt{N_2} i\frac{1}{m_{\pi\pi}}\Bigg[ (C_9\mp C_{10}) \frac{\sqrt {\lambda}}{\sqrt{ q^2}} {\cal F}_1(q^2) +2(C_{7L}-C_{7R})  \frac{\sqrt {\lambda }m_b}{\sqrt {q^2}(m_B+m_{\pi\pi})}{\cal F}_T(q^2) \Bigg],\nonumber\\
  {\cal A}_{L/R,t}^0&=&\sqrt{N_2} i\frac{1}{m_{\pi\pi}}\Bigg[ (C_9\mp C_{10}) \frac{m_B^2-m_{\pi\pi}^2}{\sqrt {q^2}} {\cal F}_0(q^2) \Bigg],\label{eq:helicity_ALR0}
\end{eqnarray}
where 
\begin{eqnarray} 
N_2= \frac{1}{16\pi^2}N_1 N_{\pi\pi} \sqrt{1-4m_{\pi}^2/m_{\pi\pi}^2},\
 N_{\pi\pi} = \sqrt{\frac{8}{3}} \frac{\sqrt {\lambda}
{q^2}\beta_\ell}{256\pi^3 m_B^3}. 
\end{eqnarray}
Here  the script $t$ denotes the time-like component of a virtual  state decays into a lepton pair.   
The function $\lambda$ is related to the magnitude of the $\pi^+\pi^-$ momentum in
$B_s$ meson rest frame: $\lambda\equiv\lambda(m^2_{B_s},m^2_{\pi^+\pi^-},
q^2)$, and
$\lambda(a^2,b^2,c^2)=(a^2-b^2-c^2)^2-4b^2c^2$.    
The combination of the time-like decay amplitude is introduced  in the
differential distribution
\begin{eqnarray}  
{\cal A}_{t}^0&=&{\cal A}_{R,t}^0-{\cal A}_{L,t}^0=  2 \sqrt{N_2}C_{10}   i\frac{1}{m_{\pi\pi}}\Bigg[  \frac{m_{B_s}^2-m_{\pi\pi}^2}{\sqrt {q^2}} {\cal F}_0(q^2) \Bigg].
\end{eqnarray}

 The $B_s\to \pi\pi$ form factors used in Eq.~\eqref{eq:helicity_ALR0} are defined by
\begin{eqnarray}
 \langle (\pi^+\pi^-)_S(p_{\pi\pi})|\bar s \gamma_\mu\gamma_5 b|\overline B_s (p_{B_s})
 \rangle  &=& -i  \frac{1}{m_{\pi\pi}} \bigg\{ \bigg[P_{\mu}
 -\frac{m_B^2-m_{\pi\pi}^2}{q^2} q_\mu \bigg] {\cal F}_{1}(m_{\pi\pi}^2, q^2) 
 +\frac{m_B^2-m_{\pi\pi}^2}{q^2} q_\mu  {\cal F}_{0}(m_{\pi\pi}^2, q^2)  \bigg\}, 
 \nonumber\\
 \langle (\pi^+\pi^-)_S(p_{\pi\pi})|\bar s \sigma_{\mu\nu} q^\nu \gamma_5 b|
 \overline B_s (p_{B_s})\rangle  &=& \frac{{\cal F}_T(m_{\pi\pi}^2, 
 q^2)}{m_{\pi\pi}(m_B+m_{\pi\pi})} \bigg[ ({m_B^2-m_{\pi\pi}^2}) q_\mu - q^2 
 P_{\mu}\bigg]. 
 \label{eq:generalized_form_factors}
\end{eqnarray}


 \begin{table}
 \caption{ The $B_s\to f_0(980)$ form factors in the
light-cone sum rules at LO and NLO in $\alpha_s$~\cite{Colangelo:2010bg}.    }
\label{tab:Bs_f0980_LCSR}
 \begin{center}
 \begin{tabular}{|l c c l | l c c l|}
 \hline 
     LO  &$F(0)$\hspace*{0.3cm} & $a_F$ \hspace*{0.3cm} &$b_F$ \hspace*{0.3cm} & NLO & $F(0)$ \hspace*{0.3cm} & $a_F$ \hspace*{0.3cm} &$b_F$   \bigstrut\\
 \hline    
  $F_1$ &$0.185\pm0.029 $ &$1.44^{+0.13}_{-0.09} $ &$0.59^{+0.07}_{-0.05}$
 &  $F_1$ &$0.238\pm0.036 $ &$1.50^{+0.13}_{-0.09} $ &$0.58^{+0.09}_{-0.07}$\\ 
  $F_0$  &$0.185\pm0.029$ &$0.47^{+0.12}_{-0.09}$ &$0.01^{+0.08}_{-0.09}$ 
  & $F_0$  &$0.238\pm0.036 $ &$0.53^{+0.14}_{-0.10}$ &$-0.36^{+0.09}_{-0.08}$   \\
 $F_T$ &$0.228\pm 0.036 $ &$1.42^{+0.13}_{-0.10}$ &$0.60^{+0.06}_{-0.05} $ 
& $F_T$ &$0.308\pm 0.049$ &$1.46^{+0.14}_{-0.10}$ &$0.58^{+0.09}_{-0.07} $
\bigstrut[b]\\
 \hline
 \end{tabular}
 \end{center}
 \end{table}

As we have shown in Ref.~\cite{Meissner:2013hya}, an explicit calculation of the $B_s\to \pi^+\pi^-$ form factors  requests the knowledge on generalised light-cone distribution amplitudes~\cite{Mueller:1998fv,Diehl:1998dk,Polyakov:1998ze,Kivel:1999sd,Diehl:2003ny,Hagler:2002nh,Pire:2008xe}.  The expressions in the light-cone sum rules  are given  as~\cite{Meissner:2013hya},
\begin{widetext}
\begin{eqnarray}
  {\cal F}_1(m_{\pi\pi}^2, q^2)&=&  N_F 
  \bigg\{\int_{u_0}^1\frac{du}{u}{\rm exp}\left[-\frac{m_b^2+u\bar u m_{\pi\pi}^2-\bar uq^2}{uM^2}\right]  \bigg[-m_b\Phi_{\pi\pi}(u)+um_{\pi\pi}\Phi_{\pi\pi}^s(u)+\frac{1}{3}m_{\pi\pi}\Phi_{\pi\pi}^\sigma(u) \nonumber\\
  &&   +\frac{
 m_b^2+q^2-u^2m_{\pi\pi}^2}{uM^2}\frac{m_{\pi\pi}\Phi_{\pi\pi}^\sigma(u)}{6}
 \bigg] 
 +\exp{[-s_0/M^2]}\frac{m_{\pi\pi}\Phi_{\pi\pi}^\sigma(u_0)}{6}\frac{m_b^2-u_0^2m_{\pi\pi}^2+q^2}
 {m_b^2+u_0^2m_{\pi\pi}^2-q^2}\bigg\},
 \label{eq:fplus} 
 \end{eqnarray}
 \end{widetext}

\begin{widetext}
\begin{eqnarray}
  {\cal F}_-(m_{\pi\pi}^2, q^2)&=&  N_F\left\{\int_{u_0}^1\frac{du}{u}{\rm
 exp}\left[-\frac{m_b^2+u\bar u m_{\pi\pi}^2-\bar uq^2}{uM^2}\right]
 \bigg[ m_b\Phi_{\pi\pi}(u)+(2-u) m_{\pi\pi}\Phi_{\pi\pi}^s(u)\right. \nonumber\\
 &&\;\;\;\left. +\frac{1-u}{3u}m_{\pi\pi}\Phi_{\pi\pi}^\sigma(u) -\frac{u({m_b^2+q^2-u^2m_{\pi\pi}^2})+2(
 m_b^2-q^2+u^2m_{\pi\pi}^2)}{u^2M^2}\frac{m_{\pi\pi}\Phi_{\pi\pi}^\sigma(u)}{6}
 \bigg]\right.\nonumber\\
 &&\left. -\frac{ u_0({m_b^2+q^2-u_0^2m_{f_0}^2})+2(
 m_b^2-q^2+u_0^2m_{\pi\pi}^2) }{u_0(m_b^2+u_0^2m_{\pi\pi}^2-q^2)}
  \exp{[-s_0/M^2]}\frac{m_{\pi\pi}\Phi_{\pi\pi}^\sigma(u_0)}{6}\right\},
  \label{eq:fminus}
  \\
 {\cal F}_0(m_{\pi\pi}^2, q^2)&=& {\cal F}_1(m_{\pi\pi}^2, q^2)+ \frac{q^2}{m_{B_s}^2- m_{\pi\pi}^2}{\cal F}_-(m_{\pi\pi}^2, q^2)\\
  {\cal F}_T(m_{\pi\pi}^2, q^2)&=&2   N_F (m_{B_s}+m_{\pi\pi}) 
 \bigg\{\int_{u_0}^1\frac{du}{u} {\rm exp}\left[-\frac{(m_b^2-\bar uq^2+u\bar
 um_{\pi\pi}^2)}{uM^2}\right]
 \left[-\frac{\Phi_{\pi\pi}(u)}{2}+m_b\frac{m_{\pi\pi}\Phi_{\pi\pi}^\sigma(u)}{6uM^2}\right] \nonumber\\
 &&   +m_b\frac{m_{\pi\pi}\Phi_{\pi\pi}^\sigma(u_0)}{6}
   \frac{\exp[-s_0/M^2]}{m_b^2-q^2+u_0^2m_{\pi\pi}^2}\bigg\},
   \label{eq:ftensor}
\end{eqnarray}\end{widetext}
where
\begin{eqnarray}
 N_F &=&B_0  F_{\pi\pi}(m_{\pi\pi}^2) \frac{m_b+m_s}{2m_{B_s}^2f_{B}} {\rm
 exp}\left[\frac{m_{B_s}^2}{M^2}\right],\nonumber\\
 u_0&=&\frac{m_{\pi\pi}^2+q^2-s_0+\sqrt{(m_{\pi\pi}^2+q^2-s_0)^2+4m_{\pi\pi}^2(m_b^2-q^2)}}{2m_{\pi\pi}^2}~.
 \label{eq:u0}
\end{eqnarray}
In the above the scalar $\pi\pi$ form factor  is defined as
\begin{eqnarray}
\langle 0| \bar ss|\pi^+\pi^-
\rangle= B_0\, F_{\pi\pi}(m_{\pi\pi}^2),
\end{eqnarray}
and the $B_0$ is the QCD condensate parameter:
\begin{eqnarray}
 \langle 0|\bar qq|0\rangle \equiv- f_{\pi}^2 B_0,
\end{eqnarray}
with $f_{\pi}$ MeV being the pion decay constant at LO. For the numerics, we use $f_{\pi}=91.4$MeV and
$ \langle 0|\bar qq|0\rangle =-(0.24\pm0.01){\rm GeV}^3$ (for a review see Ref.~\cite{Colangelo:2000dp}), which corresponds to $B_0=(1.7\pm0.2)$ GeV. 
The $M$ is a Borel parameter introduced to suppress higher twist contributions. 
Our formulae can be compared to the results for the $B_s\to f_0(980)$ transition~\cite{Colangelo:2010bg}, with the correspondence
\begin{eqnarray}
 m_{f_0}   \leftrightarrow   m_{\pi\pi}, \;\;\;\Phi_{f_0}^i(u) \leftrightarrow \Phi_{\pi\pi}^i(u),\;\;\; f_{f_0}   \leftrightarrow    B_0F_{\pi\pi}(m_{\pi\pi}^2),
\end{eqnarray}
where $f_{f_0}$ is the decay constant of $f_0(980)$ defined by the scalar current. 
The  twist-3 distribution amplitudes, $\Phi_{\pi\pi}^s(u)$ and  $\Phi_{\pi\pi}^\sigma(u)$, for the scalar $\pi\pi$ state have  the same asymptotic forms with the ones for a scalar resonance~\cite{Cheng:2005nb}, while the twist  ones can be similarly expanded in terms of the Gegenbauer moments.    Inspired by this similarity, we can plausibly  introduce an intuitive  matching:
\begin{eqnarray}
 {\cal F}_i^{B_s\to \pi\pi}(m_{\pi\pi}^2, q^2) =  \frac{1}{ f_{f_0}}  B_0F_{\pi\pi}(m_{\pi\pi}^2) F_{i}^{B_s\to f_0} (q^2).\label{eq:generalized_matching}
\end{eqnarray} 
Here we have assumed the dominance of the $f_0(980)$ which is justified in the $B_s\to \pi^+\pi^-\mu^+\mu^-$ as shown in the data in Eq.~\eqref{eq:Bs_f0_mumu_data} and Eq.~\eqref{eq:Bspipidata}.

The $B_s\to f_0(980)$ form factors  have been calculated in the light-cone sum rules at leading order (LO) and next-to-leading order (NLO) in $\alpha_s$~\cite{Colangelo:2010bg,Sun:2010nv,Han:2013zg,Wang:2014vra}, and in the perturbative QCD approach~\cite{Keum:2000ph,Keum:2000wi,Kurimoto:2001zj,Lu:2000em,Lu:2000hj,Lu:2002ny} in Ref.~\cite{Li:2008tk}. 
The momentum distribution  in the form factors has been  parametrized in the form:
\begin{eqnarray}
 F_i(q^2)= \frac{F_i(0)}{1-a_i q^2/m_{B_s^2} +b_i (q^2/m_{B_s}^2)}. 
\end{eqnarray} 
Numerical results for these quantities  where $f_{f_0}=(0.18\pm 0.015) $ GeV~\cite{DeFazio:2001uc} are taken from Ref.~\cite{Colangelo:2010bg} and  are collected in Tab.~\ref{tab:Bs_f0980_LCSR}.  Using a different value for $f_{f_0}$ for instance in Ref.~\cite{Cheng:2005nb,Cheng:2013fba} will not induce any difference to the generalized form factor, since such effects will cancel  as demonstrated  in Eq.~\eqref{eq:generalized_matching}.  In the following calculation, we will use the NLO results for the $B_s\to f_0$ transition. Using the LO results  can reduce the differential decay width by about $40\%$.

\begin{figure}\begin{center}
\includegraphics[scale=0.6]{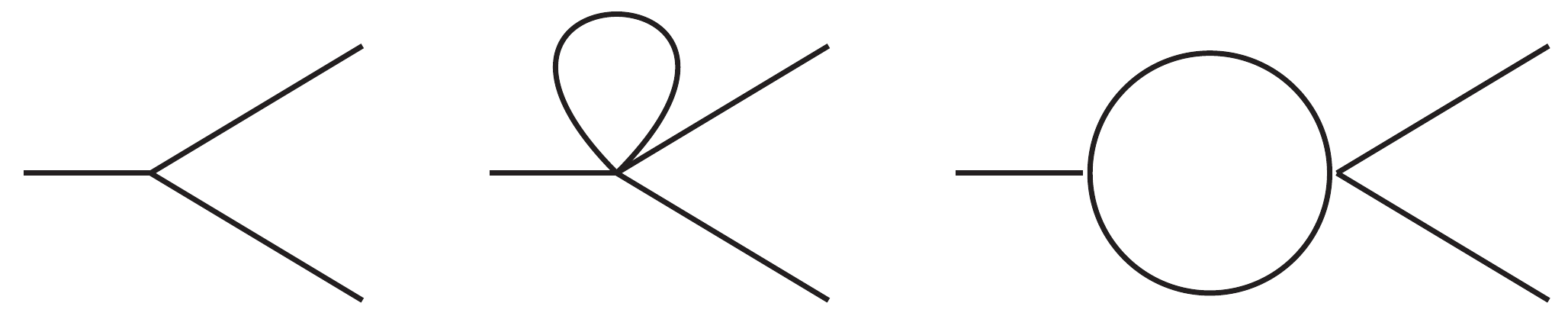} 
\caption{  Feynman diagrams for the scalar form factor at tree-level and one-loop level in CHPT. The wave function renormalization diagrams are not shown here.
} \label{fig:chpt}
\end{center}
\end{figure}

The scalar $\pi\pi$  form factor $F_{\pi\pi}(m_{\pi\pi}^2)$ has been calculated
within a variety of approaches using (unitarized) chiral perturbation
theory (CHPT)~\cite{Gasser:1990bv,Meissner:2000bc,Bijnens:2003uy,Lahde:2006wr,Guo:2012yt,Gasser:1983yg,Gasser:1984gg,Gasser:1984ux}
and dispersion
relations~\cite{Donoghue:1990xh}. In terms of  the isoscalar $S$-wave states  
\begin{eqnarray} 
|\pi\pi\rangle_{\mathrm{I=0}}^{} & =& \frac{1}{\sqrt{3}}
\left|\pi^+\pi^-\right\rangle + \frac{1}{\sqrt{6}}
\left|\pi^0\pi^0\right\rangle,\\
  |K\bar K\rangle_{\mathrm{I=0}}  &=& 
\frac{1}{\sqrt{2}}\left|K^+K^-\right\rangle + 
\frac{1}{\sqrt{2}}\left|K^0\bar K^0\right\rangle,
\end{eqnarray}
the scalar form factors  are defined as
\begin{eqnarray}
\sqrt{2}B_0\, F^{s}_1(s) &=& \langle 0| \bar ss|\pi\pi 
\rangle_{\mathrm{I=0}},  \\
\sqrt{2}B_0\,F^{s}_2(s) &=& \langle 0| \bar ss|K\bar K 
\rangle_{\mathrm{I=0}},  
\end{eqnarray}
where    the notation ($\pi$ = 1, $K$ = 2) has been introduced  for
simplicity, and the convention  $F_{\pi\pi}(m_{\pi\pi}^2)= 2/\sqrt{3} F^{s}_1(m_{\pi\pi}^2) $. In the CHPT,  expressions  have already been  derived by calculating the diagrams in Fig.~\ref{fig:chpt}  up to NLO~\cite{Gasser:1983yg,Gasser:1984gg,Gasser:1984ux,Meissner:2000bc}:
\begin{eqnarray} 
&F_1^{\rm CHPT}(s)\:\: = & \frac{\sqrt{3}}{\:2} \left[  
\frac{16 m_\pi^2}{f^2}\left(2L_6^r-L_4^r\right) + \frac{8s}{f^2} L_4^r
+ \frac{s}{2f^2} J^r_{KK}(s)
+ \frac{2}{9}\frac{m_\pi^2}{f^2} J^r_{\eta\eta}(s)
\right], \label{eq:F1CHPT}\\ 
&F_2^{\rm CHPT}(s)\:\: = & 1 
+ \frac{8 L_4^r}{f^2} \left(s - m_\pi^2 - 4 m_K^2\right)
+ \frac{4 L_5^r}{f^2} \left(s - 4 m_K^2\right)
+ \frac{16 L_6^r}{f^2} \left(4 m_K^2 + m_\pi^2\right)
+ \frac{32 L_8^r}{f^2}\,m_K^2
+ \frac{2}{3} \mu_\eta
\nonumber \\ && +
\left(\frac{9s - 8 m_K^2}{18f^2}\right) J^r_{\eta\eta}(s)
+ \frac{3s}{4f^2} J^r_{KK}(s).  \label{eq:F2CHPT}
\end{eqnarray}

\begin{figure}\begin{center}
\includegraphics[scale=0.6]{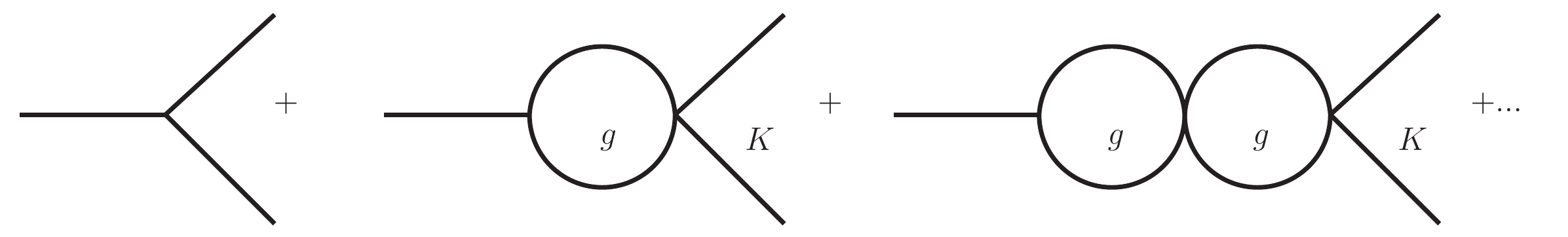} 
\caption{The  s-channel   diagrams to the scalar $\pi\pi$ form factors in CHPT. With the increase of the $\pi\pi$ invariant mass,   higher-order contributions  may become important. In the unitarized approach~\cite{Oller:1998hw}, these diagrams can be summed.  } \label{fig:uchpt}
\end{center}
\end{figure}

With the increase of the invariant mass of the $\pi\pi$ system, higher order contributions become more important. 
It has been proposed that the unitarized approach which can sum  higher order corrections and extend the applicability to the scale  around 1 GeV~\cite{Oller:1998hw}. A sketch of the resummation scheme  is shown in Fig.~\ref{fig:uchpt}.  In this figure, the $K(s)$ is the $S$-wave  projected  kernel of
meson-meson scattering amplitudes~\cite{Gasser:1983yg,Gasser:1984gg}: 
\begin{eqnarray}
&& K(s)= \left(\begin{array}{cc} K_{11} & K_{12} \\ K_{21} & K_{22} 
\end{array}\right), \\ 
&& K_{11} = \frac{2s - m_\pi^2}{2f^2}, \quad
   K_{12} = K_{21} = \frac{\sqrt{3}s}{4f^2},  \;\;\; 
   K_{22} = \frac{3s}{4f^2},
\end{eqnarray} 
where the subscript $1,2$ denotes the $\pi\pi$ and $K\bar K$  state respectively.  The function $g(s)$ is the loop integral which can be  calculated   in the 
cutoff-regularization scheme with $q_{\rm max}\sim 1$GeV being the cutoff [cf. Erratum of Ref.~\cite{Oller:1998hw}] or  in dimensional regularization. In the latter scheme,  the meson loop
function $g_{ii}(s)$ is  given by 
\begin{eqnarray}
J_{ii}^r(s) \!&\equiv&\! \frac{1}{16\pi^2}\left[
1 - \log\left(\frac{m_i^2}{\mu^2}\right) - \sigma_i(s)\log\left(
\frac{\sigma_i(s)+1}{\sigma_i(s)-1}\right)\right] = -g_{ii}(s).
\label{m_loop} 
\end{eqnarray}
with  $
\sigma_i(s) = \sqrt{1- {4m_i^2}/{s}}$.  
Imposing  the unitarity constraints, the scalar form factor can be expressed  in terms of
the algebraic  coupled-channel equation~\cite{Meissner:2000bc,Lahde:2006wr}
\begin{eqnarray}
F(s) &=& R(s) [I+g(s)K(s)]^{-1} \nonumber\\
&  =& R(s)  [I-g(s)K(s)] \:+\: \mathcal{O}(p^6), \label{G_eq}  
\end{eqnarray}
where the above equation
has been expanded up to NLO  in the  chiral expansion.   The $R(s)=(R_1(s), R_2(s))$  includes both tree-level contributions, and other  higher order corrections that have not been summed. Thus this function has no right-hand cut, and can be obtained  by matching  onto the CHPT results in Eqs.~(\ref{eq:F1CHPT},\ref{eq:F2CHPT})~\cite{Oller:1997ti,Lahde:2006wr}:  
\begin{eqnarray} 
&R_1(s)\:\: = & \frac{\sqrt{3}}{\:2} \left\{  
\frac{16 m_\pi^2}{f^2}\left(2L_6^r-L_4^r\right)
+  \frac{8s}{f^2} L_4^r - \frac{m_\pi^2}{72\pi^2 f^2}
\left[1 + \log\left(\frac{m_\eta^2}{\mu^2}\right)\right]
\right\},\\ 
&R_2(s)\:\: = & 1 
+ \frac{8 L_4^r}{f^2} \left(s - 4m_K^2 - m_\pi^2\right)
+ \frac{4 L_5^r}{f^2} \left(s - 4m_K^2\right)
+ \frac{16 L_6^r}{f^2} \left(4m_K^2 + m_\pi^2\right)
+ \frac{32 L_8^r}{f^2} m_K^2 + \frac{2}{3} \mu_\eta \nonumber \\
&& + \frac{m_K^2}{36\pi^2 f^2} 
\left[1 + \log\left(\frac{m_\eta^2}{\mu^2}\right)\right].
\end{eqnarray}

\begin{figure}\begin{center}
\includegraphics[scale=0.6]{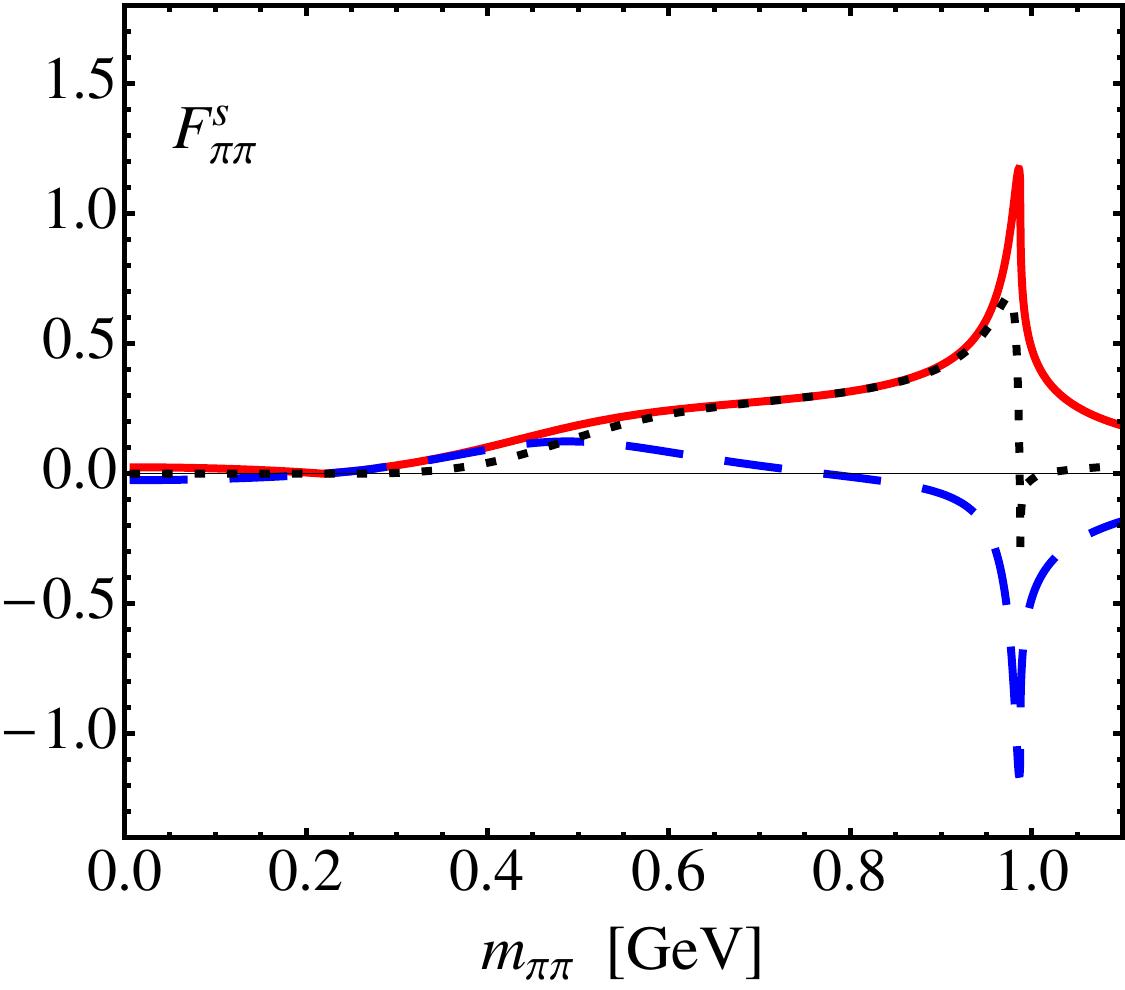} 
\caption{The  $\pi\pi$     scalar form factor   obtained in the unitarized chiral perturbation theory. The modulus, real part and imaginary part are shown in solid, dashed and dotted curves. } \label{fig:pipi_ff}
\end{center}
\end{figure}

With the above formulae and the fitted results for the low-energy constants
$L_i^r$ in Ref.~\cite{Lahde:2006wr} (evolved  from $m_\rho$ to the scale $\mu=
2q_{\rm max}/\sqrt{e}$), we show the  strange  $\pi\pi$   form factor  in Fig.~\ref{fig:pipi_ff}.  The modulus, real part and
imaginary part are shown as solid, dashed and dotted curves.

Equipped with the results for scalar form factor and heavy to light transition, we can explore  the differential branching fraction for the $B_s\to \pi^+\pi^-\mu^+\mu^-$. Our theoretical  results  for $d{\cal B}/dm_{\pi\pi}$  is given in the left panel of Fig.~\ref{fig:dBRdM2h}.   This clearly shows the peak corresponding to the $f_0(980)$.  In order to compare with the experimental data, we also give the binned results on the right panel in Fig.~\ref{fig:dBRdM2h} from $0.5$ GeV to $1.3$ GeV.  Theoretical errors shown in this panel arise from the ones in the form factors. The experimental data (with triangle markers)  has been normalized to the central value given in Eq.~\eqref{eq:Bspipidata}.   The comparison in this panel shows a general agreement  between  our theoretical prediction  and the experimental data except in a few bins. This agreement  is very encouraging.

In spite of the agreement, there exist some differences in our results and data. For instance our theoretical result  does not show the enhancement at $m_{\pi\pi}\simeq (800,1100, 1250)$MeV as given in the data. The excess at $800$ MeV may come from the tail of the $B_s\to \eta (\to \pi^+\pi^-\pi^0, \pi^+\pi^-\gamma) \mu^+\mu^-$, while in the range above 1GeV, the contribution from the $f_0(1370)$ may not be negligible.  

Integrating out the $m_{\pi\pi}$, we have the branching fraction: 
 \begin{eqnarray}
{\cal B}(B_s\to f_0(980)(\to \pi^+\pi^-) \mu^+\mu^-) = (4.1\pm 1.6)\times 10^{-8}, \label{eq:theory_Bs_f0_mumu}
 \end{eqnarray}
 which deviates from the data by about $2\sigma$. However,  one  expects  the experimental  result in Eq.~\eqref{eq:Bs_f0_mumu_data} would   get somewhat reduced.  This can be witnessed by the $B^-\to J/\psi K^-$ and $B^-\to K^-\mu^+\mu^-$~\cite{Agashe:2014kda}  
\begin{eqnarray}
\frac{
{\cal B}( B^-\to K^-\mu^+\mu^-)}{{\cal B}( B^-\to J/\psi K^-) }  =\frac{ (4.49\pm0.23)\times 10^{-7}}{ (1.027\pm 0.031)\times 10^{-3}} \sim 4.4\times 10^{-4}.
\end{eqnarray}
If this ratio were not sensitive  the light meson in the final state which is true in most cases, the branching fraction for 
the $B_s\to J/\psi f_0(980)$~\cite{Agashe:2014kda} 
\begin{eqnarray} 
{\cal B}(B_s\to J/\psi f_0)  = (1.39\pm0.14)\times 10^{-4}, \nonumber
\end{eqnarray}
would indicate 
\begin{eqnarray}
{\cal B}(B_s\to f_0(980) \mu^+\mu^-) \sim  6.1 \times 10^{-8}. 
\end{eqnarray}
This value is smaller by about $30\%$ than the central value given in Eq.~\eqref{eq:Bspipidata}, and is more consistent with our theoretical result. 
The future measurement with more data at the  experimental facilities like LHC and KEKB will be able to clarify this point, and thus to examine our theoretical formalism more precisely.   We strongly encourage our experimental colleagues to conduct such measurements. 


In summary, in this work  we have analyzed the $B_s\to \pi^+\pi^-\ell^+\ell^-$ that has focused on the region where the pion pairs  have invariant mass in the range $0.5$-$1.3$ GeV and muon pairs do not originate from a  resonance. We have adopted the approach  proposed in our previous work~\cite{Meissner:2013hya,Meissner:2013pba,Doring:2013wka} (see also Ref.~\cite{Wang:2014sba} for an overview) which makes uses of  the two-hadron light-cone distribution amplitude. The scalar $\pi^+\pi^-$ form factor induced by the strange $\bar ss$ current is predicted by the unitarized chiral perturbation theory.  The heavy to light transition can then be handled by the light-cone sum rules approach.   Merging these quantities, we have presented our theoretical results for differential decay width and compared with the experimental data. Except in a few bins, our theoretical results are in alignment with the data.  We have also discussed the disagreement and given our expectation.  More accurate measurements at the LHC and KEKB  in future are helpful  to validate/falsify  our formalism and determine the inputs in this approach.

{Acknowledgements}: The authors thank  Michael D\"oring,  Feng-Kun Guo, Bastian Kubis, Hsiang-Nan Li, Cai-Dian L\"u, Ulf-G. Meissner,  Eulogio Oset, Wen-Fei Wang for 
enlightening discussions. This work was supported in part by a key laboratory grant from the Office of Science and Technology, Shanghai
Municipal Government (No. 11DZ2260700),  and  by Shanghai Natural  Science Foundation  under Grant No.15ZR1423100.



\end{document}